 \documentclass[preprint]{aastex}

\slugcomment{}

\begin{document}


\title{Unidentified Infrared Emission\\
    Bands in the Diffuse Interstellar Medium\altaffilmark{1}}


\author{Kin-Wing Chan\altaffilmark{2,3}, T. L. Roellig\altaffilmark{2}, 
T. Onaka\altaffilmark{4}, M. Mizutani\altaffilmark{4}, 
K. Okumura\altaffilmark{5}, I. Yamamura\altaffilmark{6}, 
T. Tanab\'e\altaffilmark{7}, H. Shibai\altaffilmark{8},
T. Nakagawa\altaffilmark{6}, and H. Okuda\altaffilmark{9}}


\altaffiltext{1}{Based on observations with ISO, an ESA project with 
instruments funded by ESA members states (especially the PI countries France, 
Germany, the Netherlands, and the United Kingdom) and with the participation
of ISAS and NASA.}
\altaffiltext{2}{NASA Ames Research Center, MS 245--6, Moffett Field,
CA 94035-1000}
\altaffiltext{3}{present address:  Department of Astronomy, University of 
Tokyo, Bunkyo-ku, \\
Tokyo 113-0033, Japan,  kwc@astron.s.u-tokyo.ac.jp}
\altaffiltext{4}{Department of Astronomy, University of Tokyo, Bunkyo-ku, 
Tokyo 113-0033, Japan}
\altaffiltext{5}{Communications Research Laboratory, Koganei, Tokyo 184-8795,
Japan}
\altaffiltext{6}{Institute of Space and Astronautical Science, Sagamihara, 
Kanagawa 229-8510, Japan}
\altaffiltext{7}{Institute of Astronomy, University of Tokyo, Mitaka, 
Tokyo 181-8588, Japan}
\altaffiltext{8}{Department of Physics, Nagoya University, Chikusa-ku, 
Nagoya 464-8602, Japan}
\altaffiltext{9}{Gunma Astronomical Observatory, Gunma 377-0702, Japan}


\begin{abstract}
Using the Mid-Infrared Spectrometer on board the Infrared Telescope in
\linebreak[4]
Space and the low-resolution grating spectrometer (PHT-S) on board 
the \linebreak[4] 
Infrared Space 
Observatory, we obtained 820 mid-infrared (5 to 12 $\mu$m) spectra of the 
diffuse interstellar medium (DIM) in the Galactic center, W51, and Carina 
Nebula regions. These spectra indicate that the emission is dominated by the 
unidentified infrared (UIR) emission bands at 6.2, 7.7, 8.6, and 11.2 $\mu$m. 
The relative band intensities (6.2/7.7 $\mu$m, 8.6/7.7 $\mu$m, 
and 11.2/7.7 $\mu$m) were
derived from these spectra, and no systematic variation in these ratios was
found in our observed regions, in spite of the fact that the incident 
radiation intensity differs by a factor of 1500. Comparing
our results with the polycyclic aromatic hydrocarbons (PAHs) model for the
UIR band carriers, PAHs in the DIM have no systematic 
variation in their size distribution, 
their degree of dehydrogenation is independent of the 
strength of UV radiation field, and they are mostly
ionized. The finding that PAHs in the DIM with low UV radiation
field strength are mostly
ionized is incompatible with past theoretical studies, in which a large
fraction of neutral PAHs is predicted in this kind of environment.
A plausible resolution of this discrepancy is that the 
recombination coefficients for electron and large PAH positive ion
are by at least an order of magnitude less than 
those adopted in past theoretical studies. 
Because of the very low population of neutral state
molecules, photoelectric emission from interstellar PAHs is probably not
the dominant source of heating of the diffuse interstellar gas.
The present results imply constant physical and chemical 
properties of the carriers of the UIR emission bands in the DIM covering
the central and disk regions of the Galaxy, which could help in the
identification of the carriers.
 
\end{abstract}




\section{Introduction}
The so-called unidentified infrared (UIR) emission bands include features that 
always occur together at 3.3, 6.2, 7.7, 8.6, 11.2, and 12.7 $\mu$m. 
Since the first detection of the 11.2 $\mu$m band by Gillett, Forrest, \& 
Merrill (1973) in planetary nebulae, the UIR bands have been observed in a
variety of sources (see Tokunaga 1997 for a recent review). 
Puget, L\'eger, \& Boulanger (1985) have proposed that the excess 
\textit{IRAS} 12 $\mu$m emission
 observed in the Galactic diffuse clouds (Boulanger, Baud, \& 
van Albada 1985) can be attributed to the longer wavelength UIR emission bands. 
The detection of these emission bands in the diffuse interstellar medium (DIM)
has strongly favored this explanation of the excess \textit{IRAS} 12 $\mu$m 
emission in
the Galaxy (see Onaka et al. 1996; Mattila et al. 1996 and references therein). 
The consequences of this hypothesis imply that the carriers of the UIR bands 
play an important role in the energy balance and chemical processes in 
the Galaxy. 

To date there is still no definite identification of the carriers of the UIR 
emission bands. Several carriers have been proposed: polycyclic aromatic
hydrocarbons (PAHs; L\'eger \& Puget 1984; Allamandola, Tielens, \& 
Barker 1985), quenched carbonaceous composite (QCC; Sakata et al. 1984), 
hydrogenated amorphous carbon  (HAC; Duley \& Williams 1981), and coal grains 
(Papoular et al. 1989). Among these the PAHs are the most thoroughly studied 
in both theoretical and laboratory studies, and therefore one can make 
quantitative comparison between the observations and model calculations.
For example, one can compare the observed spectra with those of PAH spectra
measured in the laboratory. One can also compare the observed
relative UIR band intensities in different UV radiation 
environments with those predicted by the PAHs model.
In the PAHs model, changes in 
the degree of dehydrogenation and/or in the relative abundance of neutral and 
ionic PAHs can change the near- and mid-infrared band intensities
(e.g., Jourdain de Muizon, d'Hendecourt, \& Geballe 1990; Schutte, Tielens,
\& Allamandola 1993; Allamandola, Hudgins, \& Sandford 1999). 

The DIM is an ideal environment for testing the PAHs model. 
The properties of dust grains particular to their formation place 
should be averaged out and the stable species should be abundant there. 
Studies of the UIR emission bands in the 
DIM covering a wide range of UV radiation field strengths 
uniquely test the physical process operating on the interstellar PAHs.
The UIR band emission in the DIM, however, is difficult to observe with 
ground-based warm telescopes, and some bands are strongly absorbed in the
terrestrial atmosphere. High sensitivity space-borne
observations are required for this study. 
In this paper we present results from a large sample of mid-infrared 
spectrophotometeric observations in the DIM which were obtained from the 
Infrared Telescope in Space (IRTS) and the Infrared Space Observatory (ISO).
We discuss discrepancies between the present results and the predictions 
from the PAHs model.

\section{Observations}
The Mid-Infrared Spectrometer (MIRS) was one of the four science focal-plane 
instruments that flew aboard the orbiting IRTS. 
The IRTS was the first Japanese infrared space-borne telescope 
developed and operated in collaboration with NASA. It was 
launched on 1995 March 18 and surveyed $\sim$ 7$\%$ of the sky over the course
of its 26-day mission life (Murakami et al. 1996). The MIRS had an 
aperture size of 
\(8'\) $\times$ \(8'\) and operated over a wavelength range of 4.5 to 11.7 
$\mu$m with a resolution of $\Delta$$\lambda$ = 0.23 to 0.36 $\mu$m 
(Roellig et al. 1994). During 
the course of its mission, the MIRS observed part of the Galactic center and 
W51 regions and we present here the DIM data at 
${l}$ = 0$^{\circ}$ to --12$^{\circ}$ with ${|b|}$ $\leq$ 2$^{\circ}$ and 
${l}$ = 45$^{\circ}$ to 51$^{\circ}$ with ${|b|}$ $\leq$ 2$^{\circ}$. The 
uncertainty of the absolute calibration is about 10$\%$ (see Tanab\'e et 
al. 1997 for details of the MIRS calibration). The selected regions 
for this study did not include bright \textit{IRAS} point sources,
and the zodiacal background emission has been subtracted. 
To subtract the zodiacal background, we extracted background spectra 
that were observed by the MIRS at regions a few degrees of
galactic latitude and longitude away from the selected regions.
The uncertainties in the zodiacal emission subtraction in our selected 
regions are many times lower than the galactic plane emission there.   
The total far-infrared intensity, \textit{FIR}, and optical depth at 
100 $\mu$m, $\tau_{100\mu m}$, at each 
position have been obtained from Okumura (1998), who used the 155 $\mu$m 
continuum fluxes from the Far-Infrared Line Mapper (FILM) on board the IRTS 
(Okumura et al. 1996) and the \textit{IRAS} Sky Survey Atlas (ISSA; Wheelock
et al. 1994) 100 $\mu$m fluxes to 
derive those values (see also Okumura et al. 1999). The dust temperature,
\textit{T}, and $\tau_{100\mu m}$ 
were derived from a modified blackbody (a blackbody multiplied by a 
$\lambda$$^{-2}$ dust emissivity law) fit to the 100 and 155 $\mu$m fluxes. 
The \textit{FIR} is estimated as

\begin{equation}
FIR = \tau_{100\mu m} \int_{0}^{\infty} [100\mu m/\lambda(\mu m)]^{2}
B_{\lambda}(T) d\lambda ,\;
\end{equation}
where \textit{B$_{\lambda}$(T)} is the Planck function and $\lambda$ is the 
wavelength in $\mu$m. The FILM has a beam size of
\(8'\) $\times$ \(13'\)  and the ISSA 100 $\mu$m fluxes have been smoothed 
into a \(12'\) resolution with a \(4'\) bin. 
The \textit{FIR} in our selected regions has a range of 4 $\times$ 10$^{-6}$ 
to 1000 $\times$ 10$^{-6}$ W m$^{-2}$ sr$^{-1}$.
Following Okumura (1998) we estimate the incident radiation intensity 
in units of the interstellar radiation field of the solar neighborhood 
(= 1.6 $\times$ 10$^{-6}$ W m$^{-2}$, Habing 1968), \textit{G$_{0}$},
from the \textit{FIR} as

\begin{equation}
G_{0} = \frac{4\pi FIR}{1.6 \times 10^{-6}<\tau_{abs}>} ,\;
\end{equation}
where $<\tau_{abs}>$ is the absorption optical depth averaged over the incident 
radiation field. Since \textit{G$_{0}$} defined here 
covers a wavelength range from UV to visible, it should be larger 
than the \textit{G$_{0}$} defined by Habing (1968), in which the incident 
intensity includes only the UV range. 

In addition to the MIRS data, we also included data of the observations of
the Carina Nebula region with the low-resolution spectrometer (PHT-S) and
long-wavelength spectrometer (LWS) on board the ISO 
(Kessler et al. 1996). The ISO observations were made in a 
two-dimensional raster mode with a 
spacing of \(3'\) for a rectangular area of \(40'\) $\times$ \(20'\) centered 
at ${l}$ = 287$^{\circ}$.25 and ${b}$ = --0.6$^{\circ}$ (see Mizutani, Onaka,
 \& Shibai 1999 for details of the observation). The observed region 
included the DIM 
and molecular clouds in the Carina Nebula. The PHT-S had an aperture size of 
\(24''\) $\times$ \(24''\) and operated over a wavelength range of 2.5 to 11.6 
$\mu$m with a spectral resolution of about 90 (Lemke et al. 1996).
Only the data between 5.7 and 11.6 $\mu$m is used in the present study. 
The PHT-S data were reduced by PIA version 8.1.  Special care was
taken for the effects after cosmic ray hits in the
raster scan mode and the data affected by the cosmic ray hit were removed.
The LWS observations were made in the full grating scanning mode (LW01) for
 45 to 170 $\mu$m with a spectral resolution of about 200 (Clegg et al. 1996). 
The LWS beam size varies with detector,
ranging from  \(60''\) to \(84''\) FWHM (Gry 2000).
The pipeline version 7 data products were used for the LWS data.
The data were defringed by software developed based on the subroutine
in the ISO Spectral Analysis Package (ISAP\footnote{The ISO Spectral 
Analysis Package (ISAP) is a joint development by the LWS 
and SWS Instrument Teams and Data Centers. Contributing institutes are
CESR, IAS, IPAC, MPE, RAL and SRON.}). The gaps between the detectors
were corrected by scaling the fluxes relative to the SW2 detector such that
the adjacent detector signals were connected smoothly.  
All the observed positions were quite bright in the far-infrared.
Therefore the detector dark current was
insignificant and we attributed
the gaps mostly to uncertainties in the response or differences in the aperture
size of each detector. The correction does not exceed 20\% and it does not
affect the conclusions of this paper. Because the LWS spectra are
calibrated against a point-like source, a beam-size correction was
needed for observations of diffuse sources. 
The values given in Table 7.1 of the LWS Data Users Manual 
(Trams et al. 1998) were fitted to a quadratic function of the wavelength
and the beam-size correction was made based on the function. The resultant
spectra can be well fitted with a blackbody times a $\lambda^{-1}$ emissivity.
The values of \textit{FIR} and \textit{G$_{0}$} were derived in a similar 
manner as for
the ISSA/FILM data, though we used a $\lambda^{-1}$ emissivity in the LWS
data instead of a $\lambda^{-2}$ emissivity.
The beam-size correction affects the \textit{FIR} by about 30$\%$ and hence 
\textit{G$_{0}$} by the same amount.  
The absolute values
of \textit{FIR} and \textit{G$_{0}$} may have large uncertainties owing to
 this correction
and the uncertainty in the beam size (Burgdorf et al. 1997), 
while their relative scale
was less affected and the relative order was not affected at all.
The following discussions are made mostly on an order-of-magnitude basis
and are not affected by these uncertainties. 
Taking the value of $<\tau_{abs}>$/$\tau_{100\mu m}$ = 700 
(Draine \& Lee 1984), 
the range of \textit{FIR} in our selected regions yields values for 
\textit{G$_{0}$} of 3 to 4500. 
Total of 820 spectra were obtained for the following analysis. 

\section{Results}
Since the extinction towards the Galactic center and W51 regions is high even 
in mid-infrared wavelengths, extinction corrections of the spectra of
these regions were necessary. At each of the observed positions, we used the 
derived $\tau_{100\mu m}$ and then employed the extinction law of Mathis (1990)
 to correct for interstellar extinction. The extinction corrected UIR band 
ratios, which will be discussed later, show values by a few percent to 20$\%$ 
different compared to those without the extinction correction. The visual
extinction towards the Carina Nebula is low, with an average $A_{V}$ of 
$\sim$ 0.5 (Feinstein, Marraco, \& Muzzio 1973). Similar low extinction
values have been obtained from the derived $\tau_{100\mu m}$. The low 
visual extinction corresponds to less than a 1$\%$ difference between
the extinction-corrected and uncorrected UIR band ratios, and 
therefore, no extinction correction has been applied to the Carina Nebula 
spectra. 

For all of the observed positions, the MIRS and PHT-S detected four UIR bands 
at 6.2, 7.7, 8.6, and 11.2 $\mu$m. These observed mid-infrared spectra are 
similar in shape to those taken in the DIM around the W51 region (Onaka et 
al. 1996). Figure 1 shows some of the observed spectra in our three observed 
regions, and they are quite similar to each other even though
they are observed in different parts of the Galaxy. 
The intensities of the 6.2, 8.6, and 11.2 $\mu$m band relative to 
those of the 7.7 $\mu$m band have been derived. Following Onaka et al. (1996),
we drew a straight line between the edges of the emission
features for each band where the flux was assumed to come predominantly 
from the continuum and integrated the flux above the lines as a measure 
of the band intensities (see Figure 1). 
For the MIRS spectra, the wavelength points selected 
as continuum are 5.95 and 6.64 $\mu$m, 7.09 and 8.22 $\mu$m, 8.22 and 9.12 
$\mu$m, and 10.91 and 11.59 $\mu$m for the 6.2, 7.7, 8.6, and 11.2 $\mu$m 
feature bands, respectively. Similar wavelength points were selected in the 
PHT-S spectra. 

Figure 2 shows the plot of the relative band intensities of the 6.2/7.7 $\mu$m, 
8.6/7.7 $\mu$m, and 11.2/7.7 $\mu$m vs. \textit{G$_{0}$}.
No systematic variation is found in those figures. The ratios do 
not show any increasing or decreasing trend with \textit{G$_{0}$}. The
average relative band intensity ratios for the 6.2/7.7 $\mu$m, 8.6/7.7 $\mu$m, 
and 11.2/7.7 $\mu$m are 0.39, 0.24, and 0.40, respectively.




\section{Discussion}
 
The finding of no systematic variation in the UIR band ratios over a wide
range of \textit{G$_{0}$} has also been reported in the interstellar medium 
(Boulanger et al. 1998; Uchida et al. 2000). 
Uchida et al. (2000) studied a sample of reflection nebulae and found that the
UIR band ratios show no systematic variation over a range of hardness
and strength of UV radiation fields (\textit{T$_{eff}$} = 6800 to 22,000 K and 
\textit{G$_{0}$} = 40 to 1800). On the other hand, variations
of the UIR band ratio have been observed in environments covering a
wide range of physical and chemical conditions 
 (e.g., Cohen et al. 1986; Bregman et al. 1994, 1995; 
Joblin et al. 1996; Roelfsema et al. 1996; Cesarsky et al. 1996; 
Lu 1998; Mattila, Lehtinen, \& Lemke 1999).
Since the present results together with the results of Uchida et al. (2000) 
suggest that the UIR band ratios are independent of the hardness and strength
of the UV radiation field, there must be some other parameters that 
contribute to the variation in the strength of the UIR band emission in
the interstellar medium. A plausible explanation of these conflicting results 
is that the physical and chemical properties of the carriers of the UIR 
bands vary in environments where the carriers have just formed or where 
the chemical abundances are different. For PAHs model, 
the newly formed PAHs consist of a large amount of 
unstable molecules, which have very different band ratios  
compared with those of the more stable ones (Allamandola et al. 1999). 
For QCC and coal models, the available oxygen
can change the oxidization of these materials, 
which changes their emission band ratios (Sakata et al. 1987; Papoular
et al. 1989). Further studies are needed to elucidate the cause
of the variations. Because different authors
use different continuum baselines in the derivation of the 
UIR band intensity, quantitative
comparison of the derived value of the UIR band ratios in a certain
UV radiation field strength is difficult. The variations found by those
past studies could be within the range of the present data shown in Figure 2,
but they are not correlated with \textit{G$_{0}$} in the present results. 

The PAHs model we will test is 
a general model without specifying its composition. 
In the PAHs model the 6.2 and 7.7 $\mu$m bands arise from the C--C stretching 
vibrations of the PAH molecules and the 8.6 and 11.2 $\mu$m bands arise from 
the C--H in-plane and out-of-plane bending vibrations, respectively.
According to the proposed emission mechanisms, such as infrared fluorescence,
the relative UIR band ratios should not change with \textit{G$_{0}$}. 
Any change in the band ratios must be
associated with changes in the ionization, dehydrogenation, or size 
of the carriers.
For instance, neutral PAHs show quite weak 6.2, 7.7, and
8.6 $\mu$m bands compared
to the observed interstellar spectra, while these bands are enhanced 
significantly in ionized PAHs (Allamandola et al. 1999 and references therein). 
The observed PAH bands can be used to derive the physical and chemical
properties of PAH molecules in the observed astronomical environments.

The 6.2/7.7 $\mu$m band ratio displays variations among
and within sources covering a wide range of physical and chemical
properties (Cohen et al. 1986;
Bregman et al. 1995; Roelfsema et al. 1996; Lu 1998; Mattila et al. 1999).
The variations could be due to the changes of mean energy of the
incident UV photons or size of PAHs. Figure 2a shows that there is no 
systematic variation in the 6.2/7.7 $\mu$m band ratio with \textit{G$_{0}$}.
The difference in excitation energy between the 6.2 and 7.7 $\mu$m bands
is small and the variation in the spectrum of the diffuse interstellar
radiation field should not be large enough to change this band ratio. 
Therefore, PAHs in the DIM do not have a 
systematic variation in their size distribution.  
  
The 8.6/7.7 $\mu$m band ratio can be used as an indicator of the degree of
dehydrogenation of PAHs. The ratio decreases if the degree of 
dehydrogenation of the PAHs increases (e.g., Jourdain de 
Muizon et al. 1990; Schutte et al. 1993). 
By comparing the calculated results to the
observed PAH emission bands in the Orion Bar region, Schutte et al. (1993) 
concluded that the interstellar PAHs are almost fully
hydrogenated. Allamandola, Tielens, \& Barker  (1989) also concluded that,
based on the calculated hydrogen loss rates and rehydrogenation rates of PAHs,
the interstellar PAHs are fully hydrogenated if the PAHs are larger than 
25 to 30 carbon atoms. The present finding  
suggests that the degree of dehydrogenation of PAHs in the DIM
is independent of \textit{G$_{0}$} over \textit{G$_{0}$} = 3 to 4500. 

The ionization state of PAHs can be deduced from the relative 
intensities between the 7.7 and 11.2 $\mu$m band emission. 
For neutral PAHs the intensity of
the 11.2 $\mu$m band emission is higher than those of the 7.7 $\mu$m band 
emission and vice versa for ionized PAHs (Schutte et al. 1993;
Allamandola et al. 1999). A study of the spacing between the 6.2 and 
7.7 $\mu$m band emission suggested
that the interstellar PAHs have sizes of 50 to 80 carbon atoms (Hudgins
\& Allamandola 1999). Theoretical studies showed that a large fraction of 
neutral PAHs with 50 to 80 carbon atoms should exist in the DIM where 
\textit{G$_{0}$} is $\sim$ 1, and they are almost
 completely positively ionized in reflection nebulae and photodissociation
regions where \textit{G$_{0}$} is $\sim$ 10$^{5}$ 
(Omont 1986; d'Hendecourt \& L\'eger 1987; Verstraete et al. 1990; Bakes \&
 Tielens 1994; Dartois \& d'Hendecourt 1997).
Since the exact composition of PAHs existing in space is still unknown, 
it is difficult to estimate their relative band ratios in neutral and ionized
interstellar PAHs. We roughly estimate these ratios in the following by simple 
scaling. 
 
Schutte et al. (1993) found that the 11.2/7.7 $\mu$m band ratio increases 
by a factor of 5.2 when they compared this band ratio in their standard model 
with those in the neutral PAHs model.
Their standard model gave a good fit to the interstellar UIR bands
emission in the Orion Bar, where the PAHs are suggested to be totally ionized 
in a recent study by Allamandola et al. (1999). Therefore, 
we first assume that the 11.2/7.7 $\mu$m ratio increases by a factor of 5.2
when the interstellar PAHs change from 100$\%$ ionized to 100$\%$ neutral.
Then, we use the spectrum of the Orion Bar to derive the 
11.2/7.7 $\mu$m band ratio for interstellar PAHs with 100$\%$ ionized, and
finally multiply this derived band ratio by the factor 5.2 to obtain the
11.2/7.7 $\mu$m band ratio for interstellar PAHs of
100$\%$ neutral. By using the same baselines in the derivations 
of our UIR band intensities, we 
find that the Orion Bar emission shown in Allamandola et al. (1999) has
a 11.2/7.7 $\mu$m band ratio of 0.3. Therefore, the 11.2/7.7 $\mu$m band ratio
is assumed to have a value of 1.5 for 100$\%$ neutral PAHs in the 
interstellar medium. In figure 2c, almost all of the 
11.2/7.7 $\mu$m ratios are less than 1.5, which suggests that
only a small fraction of PAHs are neutral in the DIM. Assuming the
value of the band ratio can be scaled linearly between 0.3 (100$\%$ ionized)
and 1.5 (0$\%$ ionized) with the ionization fraction of interstellar PAHs, 
figure 2c suggests that the ionization fraction of PAHs in the DIM is
between 60$\%$ and 100$\%$, with a mean value of 90$\%$,
and that the degree of ionization shows no systematic variation in the 
range of \textit{G$_{0}$} = 3 to 4500. The finding here is incompatible
with those previous theoretical studies, and this
discrepancy can be resolved if the
recombination coefficients for electron-PAH ion interactions are much 
lower than those adopted in the previous theoretical studies.

The ionization fraction, \textit{f}(+), of PAHs depends on the 
photoionization rate for the neutral PAHs, \textit{R$_{ion}$}, and the 
recombination rate for electron and PAH positive ion, \textit{R$_{rec}$}, 
and can be written as

\begin{equation}
                 f(+) = \frac{R_{ion}}{R_{ion} + R_{rec}} .\;
\end{equation}
The photoionization rate for neutral PAHs is proportional to the incident
UV radiation intensity and the photoionization cross-section of the neutral 
PAHs. Based on the past laboratory measurement of the photoionization
cross-section of smaller PAH molecule, benzene (C$_{6}$H$_{6}$), and together 
with their own measurements of two larger PAH molecules, 
pyrene (C$_{16}$H$_{10}$) and coronene (C$_{24}$H$_{12}$), 
Verstraete et al. (1990) concluded that the photoionization cross-sections
of large PAH molecules (e.g., 80 carbon atoms) are proportional to their
number of carbon atoms. 
On the other hand, laboratory measurements of the recombination coefficients 
\textit{C$_{rec}$} (\textit{n$_{e}$}\textit{C$_{rec}$} = \textit{R$_{rec}$}, 
where \textit{n$_{e}$} is the electron density)
for C$_{3}$H$_{3}$$^{+}$, C$_{5}$H$_{3}$$^{+}$, C$_{6}$H$_{6}$$^{+}$, 
C$_{7}$H$_{5}$$^{+}$, and C$_{10}$H$_{8}$$^{+}$ show values ranging from
3 $\times$ 10$^{-7}$ to 10 $\times$ 10$^{-7}$ cm$^{3}$ s$^{-1}$ at 300 K,
and indicate no correlation between the measured 
coefficients and the number of carbon atoms in the ion (Abouelaziz et al. 1993).
This differs from past theoretical studies, which used classical
 electrostatics theory to adopt a coefficient that is
 proportional to the size of the PAH. For an interstellar PAH with 
50 carbon atoms at 300 K, these earlier studies would 
adopt \textit{C$_{rec}$} of about 1 $\times$ 10$^{-5}$ cm$^{3}$ s$^{-1}$ 
in their ionization fraction calculations, which is about one to two orders 
of magnitude higher than those measured by 
Abouelaziz et al. (1993) for PAHs $\leq$ 10 carbon atoms. 
The lowest UV radiation field strength in our selected regions
have \textit{G$_{0}$} = 3, 
and for a PAH of 50 carbon atoms \textit{R$_{ion}$} is calculated as
3 $\times$ 10$^{-8}$ s$^{-1}$ (see Verstraete et al. 1990 for the
calculations of \textit{R$_{ion}$}). By adopting a \textit{n$_{e}$} = 
10$^{-2}$ cm$^{-3}$
in the DIM (Omont 1986), we find from equation (3) 
that in order for PAHs to have \textit{f}(+) $\geq$ 90$\%$ (the mean value in
figure 2c), \textit{C$_{rec}$} should be less than 
3.3 $\times$ 10$^{-7}$ cm$^{3}$ s$^{-1}$.
Our estimate of \textit{C$_{rec}$} is by more than an order of magnitude 
lower than the theoretical calculations (\textit{C$_{rec}$} = 1 $\times$ 
10$^{-5}$ cm$^{3}$ s$^{-1}$), but they are quite close to the 
measured values of Abouelaziz et al. (1993). If we adopt the theoretically 
calculated \textit{C$_{rec}$} to estimate \textit{f}(+) in equation (3),
we find that \textit{f}(+) is only 23$\%$ at \textit{G$_{0}$} = 3 and is
100$\%$ at \textit{G$_{0}$} = 4500. In this situation, the 11.2/7.7 
$\mu$m band ratio will decrease from a value of 1.2 to 0.3 when the 
\textit{G$_{0}$} increase from 3 to 4500. Additional laboratory
measurements of \textit{C$_{rec}$} for larger PAHs (e.g., 50 to 100 
carbon atoms) are definitely needed to make a better estimate of the 
ionization fraction of interstellar PAHs.  

The above finding provides us useful information on the understanding
of heating of the interstellar gas. Some studies have suggested that 
interstellar PAH molecules
can be the dominant heating source of the interstellar \ion{H}{1} gas if
$\sim$ 10$\%$ of the cosmic abundance of carbon is locked up in PAHs
(d'Hendecourt \& L\'eger 1987; Verstraete et al. 1990).
In those studies the neutral PAHs absorb interstellar UV photons 
and their ejected photoelectrons heat up the surrounding 
interstellar gas. A large population of neutral PAHs is therefore required 
for maintaining the balance between heating and cooling rates.
On the other hand, another study showed that the 
interstellar PAHs with sizes less than 100 carbon atoms contribute only a 
part to the heating of the diffuse interstellar gas, and that
the major part is provided by PAHs clusters and small graphitic grains
(Bakes \& Tielens 1994). The present result suggests that interstellar PAHs 
cannot be the dominant source of heating of the diffuse interstellar gas
unless their abundance is much larger than those adopted in the
past studies.

Finally, the present finding of no
systematic variation in the UIR band ratios in the DIM covering the
central and disk regions of the Galaxy over a wide range of UV radiation field
strengths provides a new piece of information for the identification 
of the carriers of the UIR bands. In the PAHs model, the
fitting of the observed UIR band emission in different astronomical 
environments requires a mixture of 
PAH molecules with different sizes and thermodynamical stability
(Allamandola et al. 1999). Our finding suggests that the size distribution
of the PAHs is the same in all locations and their 
thermodynamical stability is high. In the QCC and coal models, 
the enhancement of the 6.2, 7.7, and 8.6 $\mu$m bands has been 
attributed to oxidization (Sakata et al. 1987; Papoular et al. 1989). 
Therefore, the constant band ratio of 11.2/7.7 $\mu$m suggests that these 
species must be chemically stable in the DIM.

\section{Summary}

The major result of this paper is that there is
 no systematic variation in the UIR 
6.2/7.7 $\mu$m, 8.6/7.7 $\mu$m, and 11.2/7.7 $\mu$m band ratios in the DIM
covering the central and disk regions of the Galaxy over 
a wide range of UV radiation field strengths 
(\textit{G$_{0}$} = 3 to 4500).
Comparing our results with the PAHs model:  
(1) the 6.2/7.7 $\mu$m band ratio suggests that
PAHs in those environments have no systematic variation in their
size distribution; (2) the 8.6/7.7 $\mu$m band ratio suggests that the 
degree of dehydrogenation of PAHs is independent of the strength of UV
radiation field over a range of \textit{G$_{0}$} = 3 to 4500;
and (3) the 11.2/7.7 $\mu$m band ratio suggests that the PAHs are 
mostly ionized in the DIM even though the UV radiation field strength
is low. The last finding is incompatible with past theoretical studies,
in which a large fraction of neutral PAHs is predicted in this kind of 
environment. Laboratory measurements of the recombination coefficients 
of large PAHs are needed to make a better estimate of the ionization 
fraction of interstellar PAHs.
The finding that interstellar PAHs have very low
abundance of neutral state molecules in the DIM does not support the 
suggestion that PAH molecules are the dominant source of heating of the 
diffuse interstellar gas by photoelectric emission.
Finally, the present results imply that the carriers of the UIR bands
have constant physical and chemical properties in the DIM, and this 
new information could help in the identification of the carriers.

\acknowledgments
The ISOPHOT data presented in this paper were reduced using PIA, which is a
joint development by the ESA Astrophysics Division and the ISOPHOT Consortium
with the collaboration of the Infrared Processing and Analysis Center (IPAC).
Contributing ISOPHOT Consortium institutes are DIAS, RAL, AIP, MPIK, and MPIA. 
We are grateful to Louis Allamandola, Douglas Hudgins, and Jesse Bregman for 
helpful discussions about interstellar PAHs. We also thank IPAC for their help 
in the pointing reconstruction of the IRTS and the entire IRTS team for their 
efforts in ensuring the success of the IRTS mission. We also thank K. Kawara, 
Y. Satoh, and the Japanese ISO team for their continuous help and 
encouragement. We thank M. Burgdorf for the latest information on the LWS
beam and the LWS IDT in Rutherford Appleton Laboratory, particularly
 S. Sidher, for the LWS data reduction. 
This work was performed while K. W. C. held a National 
Research Council-(NASA Ames Research Center) Research Associateship, and was
supported by NASA Grant 399-28-01. K. W. C is currently supported by the
JSPS Postdoctoral Fellowship for Foreign Researchers. This work was supported 
in part by Grants-in-Aid for Scientific Research from JSPS.





\clearpage



\figcaption{Samples of observed MIRS spectra at the (a) W51 (${l}$ = 
48$^{\circ}$.75, ${b}$ = 0$^{\circ}$.00) and (b) Galactic center (${l}$ = 
--4$^{\circ}$.20, ${b}$ = 0$^{\circ}$.13) regions, and PHT-S spectra at the
(c) Carina Nebula (${l}$ = 
287$^{\circ}$.305, ${b}$ = --0$^{\circ}$.686) region. 
The dashed lines show the continuum points used in the derivation 
of the UIR band ratios.
In (c) the UIR bands are sharper 
than those in (a) and (b) and this is due to the high spectral 
resolutions of the PHT-S compared to those of the MIRS. \label{fig1}}

\figcaption{Relative band intensities of the UIR (a) 6.2/7.7
$\mu$m, (b) 8.6/7.7 $\mu$m, and (c) 11.2/7.7 $\mu$m vs. \textit{G$_{0}$}. 
The diamonds, plus symbols, and open squares are the data for the W51, 
Galactic center, and Carina Nebula regions, respectively. The scatter
in the figures are largely due to statistical errors of the 
data, and the large scatter in (b) arises from the weak strength
of the 8.6 $\mu$m band. In (c) the dashed lines at ratio of 0.3 and 1.5
 mark the expected PAHs band ratio of 100$\%$ ionized and 100$\%$ neutral,
respectively. See text for details of the derivation of these 
 ratios. \label{fig2}}


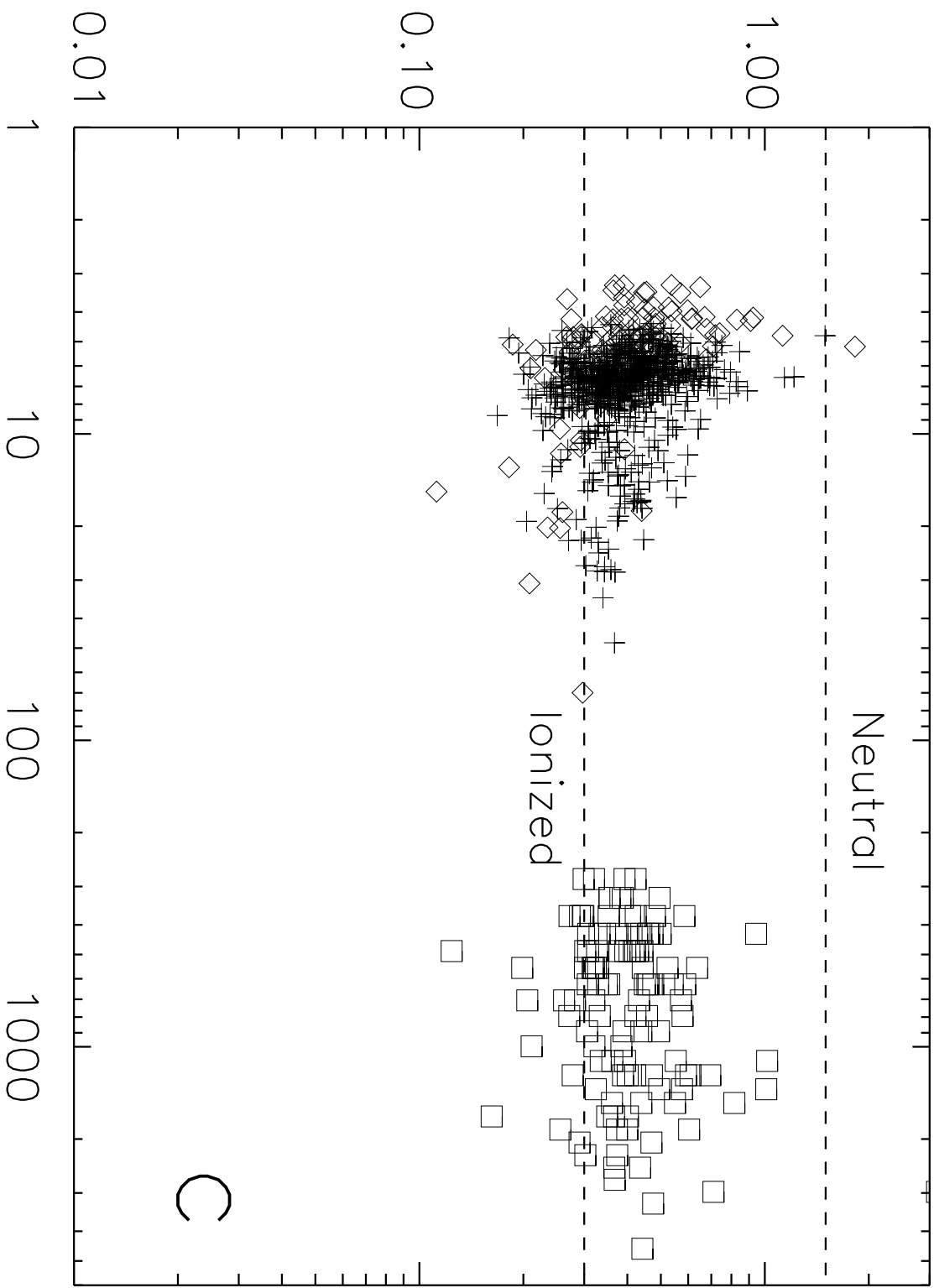
\end{document}